\documentclass[prb, floatfix, preprint]{revtex4}
\usepackage{graphicx}

\newcommand{\eq}[1]{Eq.~\ref{#1}}

\newcommand{\fig}[1]{Fig.~\ref{#1}}

\newcommand{\sect}[1]{Section\ \ref{#1}}

\begin{document}

\title{Small dimensional microstrips embedded with ferromagnetic layers: Numerical simulations and experimental results}

\author{Jonah N. Gollub${}^1$} 
\author{Bijoy Kuanr${}^2$}
\author{Zbigniew Celinski${}^2$}
\author{ Robert Camley${}^2$}
\author{David R. Smith${}^3$}

\affiliation{${}^1$ Department of Physics, University of California,
San Diego, CA, 92037 USA}
\affiliation{${}^2$ Department of Physics, University of Colorado, Colorado Springs, CO, 80918 USA}
\affiliation{${}^3$ Department of Electrical and Computer
Engineering, Duke University, Durham, NC, 27708 USA}

%\affiliation{J. Gollub is with the Univesity of California, San Diego}
%\thanks{B. Kuarn, Z. Celinski, R. Camley are with the University of Colorado, Colorado Springs.}
%\thanks{D. Smith is with Duke University}}

%\IEEEpeerreviewmaketitle

%\author{Jonah N. Gollub$^{1}$}
%\author{Bijoy Kuanr$^{2}$}
%\author{Zbigniwiev Celinski$^{2}$}
%\author{Robert Camley$^{2}$}
%\author{David R. Smith$^{1,3}$}
%\affiliation{${}^1$ Department of Physics, University of California,
%San Diego, CA, 92037 USA}

%\affiliation{${}^2$ Department of Physics, University of Colorado,
%Colorado Springs, CO, 80918 USA}

%\affiliation{${}^3$ Department of Electrical and Computer
%Engineering, Duke University, Durham, NC, 27708 USA}

%\date{\today}

\begin{abstract}
We use a numerical electromagnetic simulation software to investigate a filtering device consisting of a small dimensional microstrips embedded with a thin layer of ferromagnetic material and we compare our results to experimental results. We are able to show good correlation of simulation versus experiment for the magnitude of insertion loss and phase shift. The microstrips considered have dimensions on the order of the skin depth of the conductor and hence the field distribution is not easily calculated by analytic methods. We show that numerical simulation methods provide an accurate means of characterizing these structures.   
\end{abstract}

\pacs{Valid PACS appear here} 
\maketitle

%\preprint{APS/123-QED}

\section{\label{IntroTheory} Introduction}
The advances in communication technology have generated interest in the development of smaller and higher frequency microwave devices.  Currently under consideration for such applications are devices that incorporate high magnetic saturation materials directly into miniaturized microstrip structures. These structures are advantageous for their broad frequency range of operation, economical fabrication, and ease with which they can be integrated into microstrip based monolithic circuits. In this paper we investigate a microstrip based structure, shown in \fig{SimStruct}, that operates as a notch filter. The filter, proposed by Schoelmann et al., is formed by embedding a thin ferromagnetic layer at some position between the top and bottom conductors in a microstrip \cite{Schloemann1988, Liau1991, Cramer2000}. The thin ferromagnetic layer has a resonance governed by the equation 
\begin{equation}
\omega=\mu_0\gamma\sqrt{H_0(H_0+M_0)}, \label{FMR_eq}
\end{equation}
where $\gamma$ is the gyromagnetic ratio, $H_0$ is the applied static field, and $M_0$ is the saturation magnetization. For materials such as iron or permalloy (Py) with high magnetic saturation values, $M_0$, a modest magnetic biasing field, 0-2 kG, gives a ferromagnetic resonance above 20 Ghz.  At the resonant frequency, an embedded ferromagnetic layer in a microstrip will absorb power from the propagating quasi-TEM mode. By altering the intensity of the applied static magnetic field, $H_0$, one can tune the position of this absorption. 

While the position of maximum attenuation is predicted by \eq{FMR_eq}, solving for the magnitude of insertion loss or phase shift analytically is intractable for small dimensional microstrips where the dimensions are much smaller than the propagation wavelength. There exists an analytic solution for the related parallel plate waveguide structure embedded with a ferromagnetic layer, by Camely et al.  \cite{Camley:waveguideAnalytic,Astalos1998}, but numerical simulations are necessary to characterize the more complicated fields and current distributions present in microstrip structures. In this paper we investigate these structures using electromagnetic numerical simulation software and compare our results to experimental work.

The structure of this paper will consist of a discussion of the setup of our numerical simulations, a discussion of simulations of small dimensional microstrips and comparison to theory, and a discussion of simulations of ferromagnetically embedded microstrips and comparison to experimental results. 

\section{Numerical simulations}

\begin{figure}[t]
\centering
\includegraphics[scale=1]{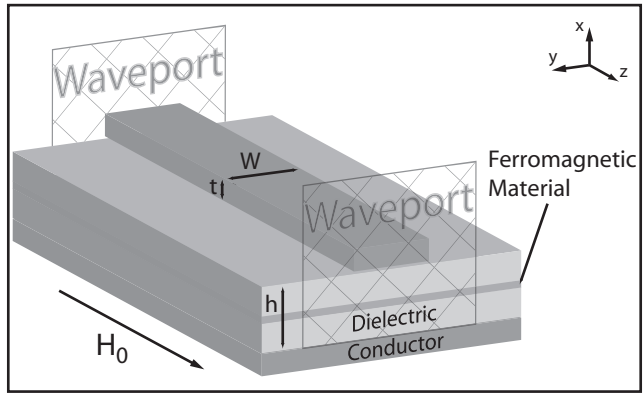}
\caption{Diagram for the setup of the ferromagnetically embedded microstrip structure. Note that the ferromagnetic layer does not actually touch the waveport because HFSS does not allow a material with a negative permeability to  be in direct contact with the waveports. An infinitesimal spacer of dielectric ($\sim$1/100000 of the effective wavelength) is placed between the waveport and ferromagnetic layer to avoid simulation issues.} \label{SimStruct}
\end{figure}

To simulate microstrip structures, we used Ansoft's commercially available electromagnetic simulation software (HFSS) which is a finite element solver. Several software packages have been optimized for microstrip structures, including \cite{Heinrich1997,Heinrich1999}, but HFSS allowed us the ability to easily integrate anisotropic materials without modification of the software.  When the dimensions of the microstrip are significantly smaller than the propagation wavelength and on the scale of the skin depth of the conductors, the magnetic and electric fields vary strongly throughout the structure. Further complicating the field distribution is the introduction of a sub-micron layer of ferromagnetic material which is both conductive and strongly magnetic. HFSS has an efficient meshing algorithm which is crucial to minimizing the solution time necessary to solve these structures, but even with optimized meshing, densely meshed models are necessary to converge on accurate results.  

Illustrations of the simulation and waveport setup are shown in \fig{SimStruct}.  The width of the simulated volume was set to $20*W$ and the height of the bounding volume was set to $10*h$, where $W$ is the width of the top strip and $h$ is the height of the dielectric. These dimensions provided a large enough simulation volume such that the propagating fields are minimal near the boundaries and do not exit the model (as was checked by plotting the fields). The boundaries of the simulation volume were set to be radiative which absorb outgoing radiative modes, but as noted above, the power exiting these surfaces is insignificant.

The length of the microstrip simulated was set to minimize simulation volume but still provide accurate simulation results. HFSS is effective at decoupling the microstrip mode from any evanescent fields extending from waveport  1 to waveport 2. We were able to simulate a microstrip with a length of as little as 1/300 of the effective microstrip wavelength (or .03 mm at 20 Ghz) with results that accurately compared to within 5\%  of the results of a microstrip 100 times as long. 

The process of meshing the simulated structure is done through a series of adaptive passes until the S-parameter solutions of the waveports change by less than .01\% between passes. A discrete sweep, which calculates the fields of the model at each frequency requested, was performed over the frequency range of interest. We investigated using HFSS's fast sweep option which interpolates results from calculations at one frequency, hence reducing simulation time, but we found that it did not give accurate results when the frequency dispersive ferromagnetic layer was included in the microstrip. The ferromagnetic layer can have a permeability where the real part ranges from -100 to 100 near resonance.  The ferromagnetic permeability was calculated in MATLAB using \eq{FMRdiagonal} and \eq{FMRoffdiagonal}, as explained in \sect{FMRsection}, and then imported as a material parameter into HFSS.

The microstrip structure is excited in the simulation by waveports outlined on rectangular surfaces at both ends of the microstrip structure as shown in \fig{SimStruct}.  HFSS uses the boundary conditions of the model touching the 2D waveport surface to determine a waveguide geometry (perfect electric boundaries are assumed on the edges of the waveport unless otherwise defined). From this geometry the possible propagation modes are calculated and used to excite the structure. For the waveports touching the microstrip structure, as shown in \fig{SimStruct}, this equates to a rectangular coaxial waveguide (partially filled with a dielectric material) feeding the microstrip structure.  Exciting the fundamental TEM mode of the waveport couples to the quasi-TEM mode of the microstrip. For these simulations we set the dimensions of the waveport to have a height of $6*h$ and width of $5*W$. 

HFFS calculates the scattering matrix by exciting the fundamental TEM mode of the coaxial waveport normalized to the input impedance of the microstrip structure touching the waveport. At the opposite waveport HFSS decouples the incident quasi-TEM mode of the microstrip back into the fundamental TEM mode of the waveport, and again normalizes by the impedance of the microstrip structure. The normalization by the input and output impedance of the structure ensures that the scattering parameters of the microstrip structure are properly de-embedded from the two coaxial waveports. One technical issue in this excitation method is that the ferromagnetic material cannot touch the waveport. Ferromagnetic materials have negative permeability values which prevent a proper solution of the waveport modes and therefore are not allowed to touch the waveports by HFSS. In our simulations we place an infinitesimal amount of dielectric material ($\sim$1/100000 of the effective wavelength of interest) between the ferromagnetic material and the waveports to avoid this error. This spacing material is much smaller than the meshing elements and does not significantly affect the simulation.  

Once the scattering parameters are derived for a short structure, it is possible to obtain those for a longer uniform structure (for comparison to experimental structures) using standard conversion formulas  \cite{Pozar1998book}. The conversion involves first converting the scattering parameters matrix, $ S$, to a transmission matrix, $A$. The transmission matrix can then be decoupled into an eigenvector matrix, $V$, and a diagonal eigenvalue matrix, $D$.
\begin{equation}
A=VDV^{-1}.
\end{equation}
The eigenvalue matrix, $D$, contains the propagation constant information, $e^{\pm\gamma}$, such that raising $D$ to the $n$th power provides the phase shift due to the propagation delay through a device $n$ times the length of the original device (where $n$ does not have to be an integer). To get the transmission matrix for a desired length, $l$, in terms of the original simulation length, $l_s$, we solve
\begin{equation}
A(l)=VD^{l/l_s}V^{-1}.
\end{equation}
A standard conversion of the transmission matrix, $A(l)$, back to a scattering parameter matrix, $S(l)$, provides the final desired scattering matrix. 

When experimental results are compared to simulation it is often the case that the constructed samples are not perfectly matched to the feed line that they are tested with, i.e. 50 Ohms. HFFS can be used to solve for the input impedance of the structure and any mismatch with a 50 Ohms line can be adjusted for in the translation $S\rightarrow S(l)$. These data manipulations were implemented using a Matlab script. 

Matlab was also used extensively in running the simulations.  A freeware `HFSS-MatLab Scripting-API' toolbox written by V.\ C.\ Vijay was used to write scripts which produced the HFSS microstrip model and assigned material parameters. This script based approach allowed easy adjustment of simulation parameters. 

\section{Small Dimensional Microstrips\label{BasicMicro}}

At microwave frequencies, the insertion loss of a microstrip is predominantly due to ohmic losses in the imperfect conductors. With knowledge of how the current is distributed over the conductors, one can integrate Joule's law over the cross-section of the microstrip to determine the attenuation constant. This is easily done for a microstrip with a wide top strip, i.e. $W\gg h$ because the current is uniformly distributed over the bottom of the top strip and the top of the ground plate. The attenuation constant is found to be 
\begin{equation}
\alpha_c\approx 8.68 /(\sigma \delta Z_0 w)\,\, {\rm dB/cm} 
\end{equation}
where $\sigma$ is the conductivity, $\delta=1/\sqrt{\pi  \mu \sigma f}$ is the skin depth, and $Z_0$ is the characteristic impedance of the structure \cite{Pucel1968}.  But, as Pucel et al.\ demonstrate \cite{Pucel1968,Pucel1968a}, this approximation grossly overestimates the losses when $W/h<2 $ because the current density over the top strip and ground plate are no longer uniform. In fact, because of the mathematical complexities involved \cite{Cockcroft1929} an exact analytical solution has never been found. Pucel et al.\  calculated an approximate analytic expression for the associated losses based on Wheeler's conformal mapping technique which is valid for small $W/h$ values and conductors with a thickness of at least four skin depths. Beyond this limit, direct electromagnetic simulation may be the best method to determine losses.

\begin{figure}[t]
\centering
\includegraphics[scale=1]{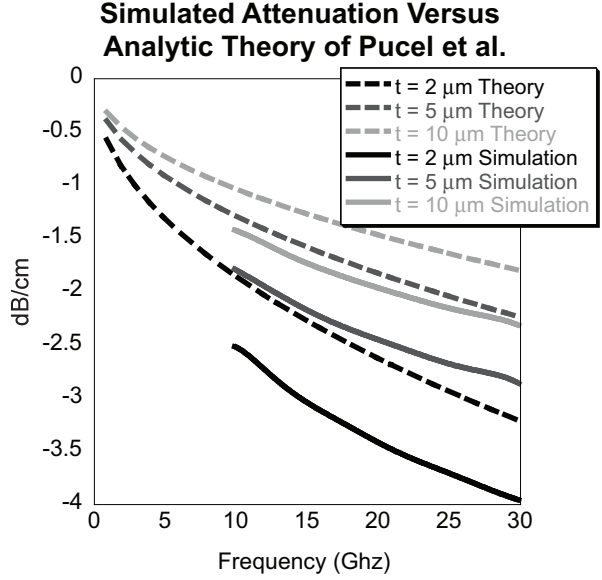}
\caption{A comparison between the attenuation derived from an HFSS simulation and that of the approximate analytic solution by Pucel et al.\ is shown for a microstrip with a dielectric height of $h=4.5\ \mu$m and a top strip width of $W=5\ \mu$m and thickness of $t=$2, 5, and 10 $\ \mu$m. The conductor is silver ($\sigma=6.1\times10^7\,$S/m) and the dielectric is silicon dioxide ($\epsilon_r=4$) with zero losses. } \label{ConductorThickness}
\end{figure}

As a precursor to simulating ferromagnetic microstrip structures we first verified our technique against Pucel's theory for small dimensional microstrip structures in the realm where $W$, $h$, and $t$ are on the order of the skin depth. We simulated a microstrip structure as shown in \fig{SimStruct}, but without the ferromagnetic layer. The parameters were $W= 5\,\mu$m, $h=4.5\,\mu$m, $t=2\,\mu$m and with silver as the conductor and lossless silicon dioxide as the dielectric ($\epsilon_r=4$).  For reference the skin depth of silver is  0.4-0.7 $\mu$m between 10-30 Ghz.  The results are shown in \fig{ConductorThickness} versus the theory of Pucel et al. and it can be seen that our simulations predict a loss of about 20\% greater than their theory. This is in line with the higher losses shown in Pucel's own experiments \cite{Pucel1968}.  It can also be seen that for both theory and simulation, the attenuation is very sensitive to the thickness of the top strip.  Further simulations, not shown, found a much smaller variation in attenuation from changing the top strip width for $t>\delta$. 

\section{Microstrips with embedded ferromagnetic layers \label{FMRsection}}

The permeability of a ferromagnetic material biased along the $z$-axis is generally described by an anisotropic tensor with off-diagonal components
\begin{equation}
\mu=
\mu_0\left(\begin{array}{ccc}
\mu_1 & i \mu_2 & 0\\
-i\mu_2&\mu_1&0\\
0&0&1\\
\end{array}\right),\label{FMRmatrix}
\end{equation}
where,
\begin{eqnarray}
\mu_1&=&1+\frac{\mu_0\gamma M_0(\mu_0\gamma H_0-i \Gamma\omega)}{(\mu_0\gamma H_0-i\Gamma \omega)^2-\omega^2},\label{FMRdiagonal}\\
\mu_2&=&\frac{\mu_0\gamma M_0 \omega}{(\mu_0\gamma H_0-i\Gamma \omega)^2-\omega^2}.\label{FMRoffdiagonal}
\end{eqnarray}
In these equations $H_0$ is the applied field, $M_0$ is the saturation magnetization, $\gamma$  the gyromagnetic ratio, $\Gamma=\gamma\Delta H/(1.16\omega)$ is the damping factor ($\Delta H$ is the damping linewidth), and $\omega$ is the angular frequency of operation. 

For a thin layer of ferromagnetic material incorporated into a parallel waveguide structure, Schloemann et al.\ \cite{Liau1991,Schloemann1988} suggested that it is possible to approximate the permeability tensor in a simpler diagonal tensor. The Schloemann approximation is given by
\begin{equation}
\mu=
\mu_0\left(\begin{array}{ccc}
\frac{\mu_1^2-\mu_2^2}{\mu_1} & 0&0\\
0&\frac{\mu_1^2-\mu_2^2}{\mu_1}&0\\
0&0&1\\
\end{array}\right).\label{SchloemannApprox}
\end{equation}
and is derived by assuming the resulting propagating mode in the structure remains substantially TEM in nature.  This has been verified rigorously by Astalos and Camley \cite{Astalos1998}. 

Schloemann maintains that this should also hold for a microstrip structures whose modes are strongly TEM in nature over the central region of the microstrip where the ferromagnetic layer is most strongly excited \cite{Schloemann1988}.   The Schloemann approximation greatly simplifies the complexity of the problem and agrees with experiment results. We use this approximation in our simulations because it simplifies the simulation and resolves the technical issue that HFSS does not directly support material parameters with off-diagonal components.  

Using \eq{SchloemannApprox} the $\mu_{xx}$ and $\mu_{yy}$ components of the permeability were calculated in MATLAB and imported into HFSS as a data table of real values (Real[$\mu$]) and loss tangent values (Imag[$\mu$]/Real[$\mu$]).

\begin{figure}[t]
\centering
\includegraphics[scale=1]{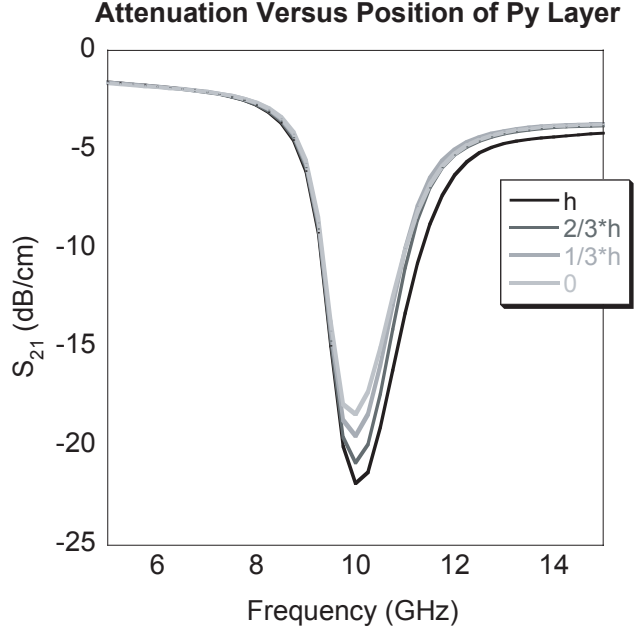}
\caption{The simulated $S_{21}$ attenuation is plotted for various positions of a ferromagnetic permalloy layer as measured from the ground plate. The  dielectric is silicon dioxide $_2$ ($\epsilon=4$). The permalloy is 0.1 $\mu$m thick and has a width extending to the edges of the simulation (100 $\mu$m). The dimensions of the microstrip are $h=3.1\ \mu$m, $W=5\ \mu$m, and $t=1.5\ \mu$m. The parameters for permalloy are $\sigma=1.6\times 10^6\ $S/m, $M_0=0.8\ $kG, and linewidth of $0.08\ $kG. The applied field is set to 1 kG with a resulting resonance near 10 Ghz.} \label{PyPosition}
\end{figure}
We considered how the position of a ferromagnetic layer affects attenuation.  For a microstrip, the magnetic field circulates around the top strip and hence provides a greater magnetic flux through a ferromagnetic layer that is closer to the top strip. Placement of a ferromagnetic layer directly under the top strip will result in the strongest magnetic excitation of the material and strongest attenuation at ferromagnetic resonance. This was verified by our simulations shown in \fig{PyPosition}, in which a thin layer of 100 nm permalloy layer was placed at various heights above the ground plate. It is shown that a layer directly below the top strip has a 4 dB/cm larger insertion loss than the one directly above the ground plate. Alternatively, for a parallel waveguide structure the attenuation is independent of layer position since the magnetic flux is constant for all positions. This was verified by simulation but is not shown here.

\section {Experimental Results and Simulation}
\begin{figure}[t]
\centering
\includegraphics[scale=1]{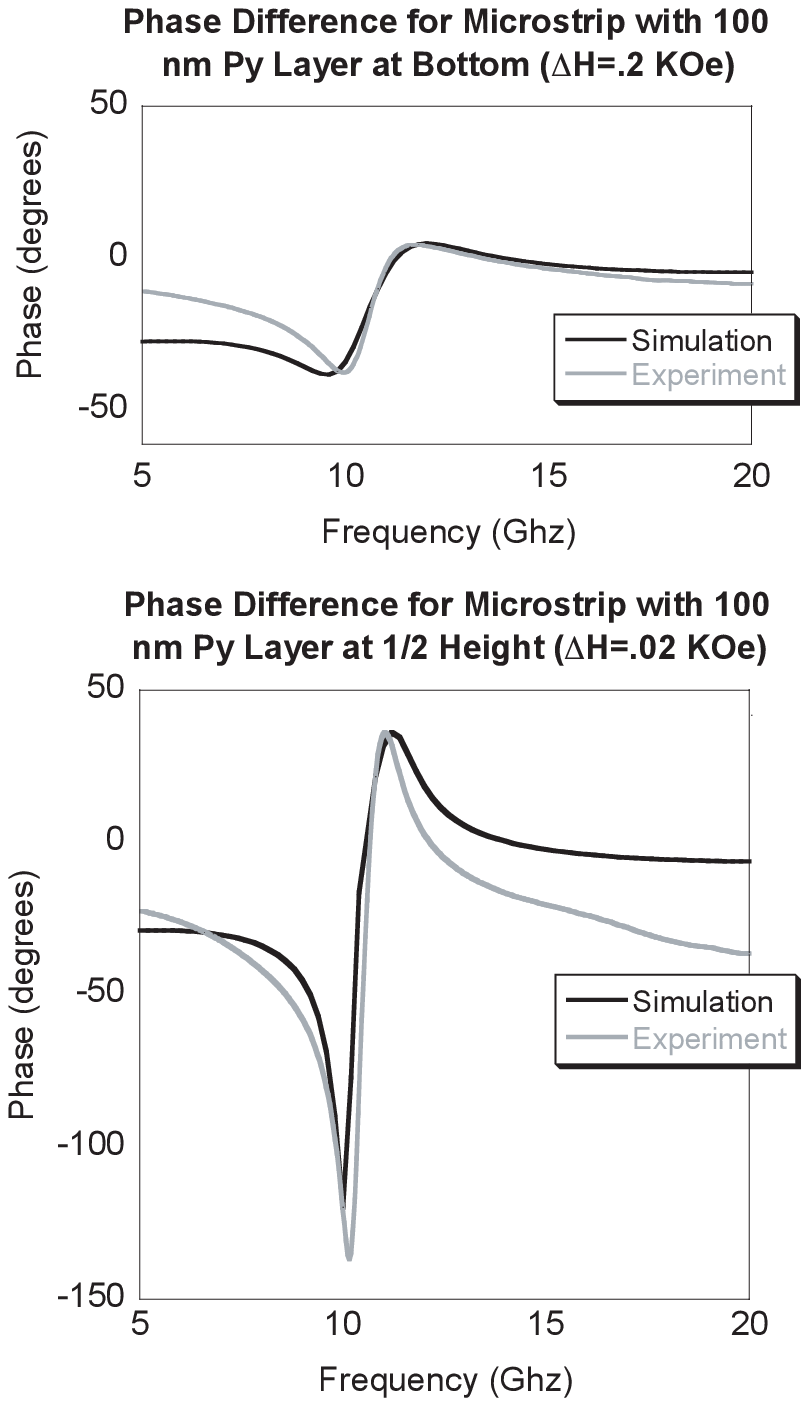}
\caption{The simulated phase difference is plotted for two positions of a ferromagnetic permalloy layer: directly above the ground plate and at half-height of the microstrip.  The conductor was copper, the dielectric was silicon dioxide ($\epsilon=4$) and  the permalloy layer was $0.1\ \mu$m thick. The dimensions of the microstrip were $h=3.1\ \mu$m, $W=5\ \mu$m, and $t=1.5\ \mu$m. The parameters for permalloy used in the simulation were $\sigma=1.6\times 10^6\ $S/m, $M_0=0.8\ $KG. The experiment was performed with a magnetic biasing field of $H_0=1$ kOe; in the simulation $H_0=1.1$ kOe to fit the data. The linewidth parameter in the simulation, $\Delta H$, was adjusted to fit the phase at resonance as given above. The simulated data was shifted down 30 degrees to match the experimental data.} \label{PyPhase}
\end{figure}
\begin{figure}[t]
\centering
\includegraphics[scale=1]{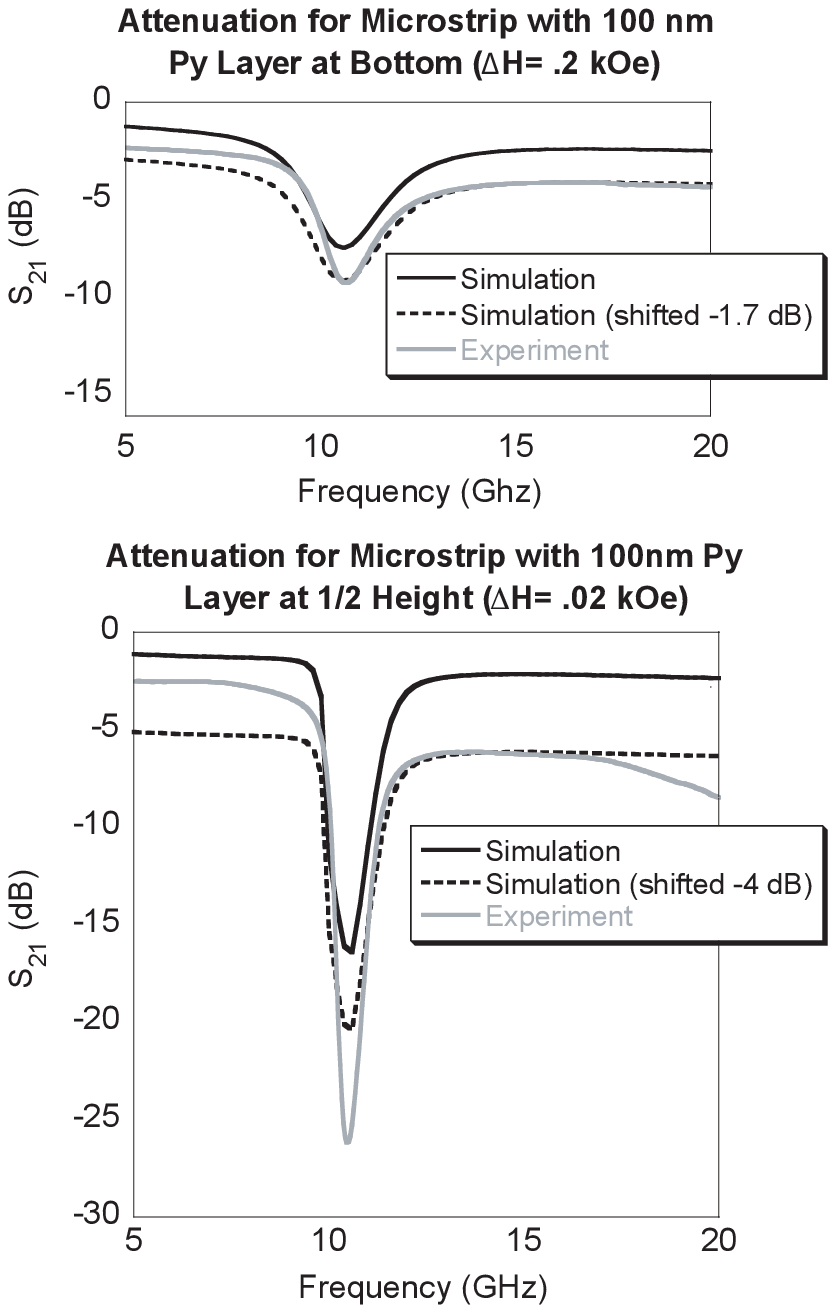}
\caption{The experimental and simulated attenuation corresponding to \fig{PyPhase}.} \label{PyMag}
\end{figure}
We further compared our simulations to experimental results. The filter structure was fabricated on a 0.5 mm silicon substrate.  The structure was grown in a sputtering system with a background pressure maintained at $\sim$1x10-7 Torr. After cleaning the silicon substrate in an ultrasonic bath, we annealed it to $300^\circ$ C inside the vacuum chamber. All the depositions were done at room temperature. First a 5 nm thick Ta layer was deposited to achieve good adhesion of the grown structure to the substrate.  This was followed by a 2 microns thick copper layer, which was used as the ground plane for our device. The next sequence of depositions, silicon dioxide (total thickness = 3 $\mu$m), Permalloy (100 nm), and finally 2 microns of copper, were made through a shadow mask.  All depositions were carried out using magnetron sputtering.  We patterned the film by photolithography and then dry etched to obtain the required strip widths and lengths for the devices.We used a variety of positions for the permalloy layer, as is discussed in the text. 

The filter characterization was done using a vector network analyzer along with a micro-probe station in the frequency range from 100 MHz to 40 GHz and under static magnetic fields up to 4.5 kOe along the long axis of the microstrip line.  Using a standardized copper coplaner structure (non-magnetic) obtained from the National Institute of Standards and Technology (NIST) at Boulder and NIST's Multical\textregistered\  calibration software, \cite{Marks1991}, a through-short-line (TRL) calibration was performed. 

The microstrip operates in a TM mode, which ensures the ferromagnetic resonance condition because the RF magnetic field and the DC magnetic field are perpendicular to each other. The signal lines in our structures have a width (W) of 5 $\mu$m, and a lengths (L) of 6 mm. The microstrips were generally designed to be close to a 50 $\Omega$ characteristic impedance. For each filter the magnitude of S-parameters and phase advance were measured. 

In the samples constructed, the magnetic quality of the permalloy film is dependent on the manufacturing process. A layer that is grown direcltly above the ground plate may have a substantially different linewidth compared to one grown at an intermediary position between the silicon dioxide layers. The linewidth of a permalloy film in the microstrip is difficult to measure directly but we can estimate it from simulations by using it as a fit parameter. In \fig{PyPhase}, the simulated versus experimental phase difference of $S_{21}$ is plotted for two structures, one with the permalloy layer directly above the ground plate and another with the permalloy layer at the half-height position. The phase difference plot compares the phase difference between a microstrip with a 100 nm permalloy embedded layer and a microstrip with identical dimensions but no permalloy layer. In the simulations the linewidth parameter was varied until it fit the phase shift near resonance (note that the simulations were shifted down 30 degree to correlate with the experimental data at resonance). It can be seen from the fitted values of $\Delta H$ that the quality of the permalloy layer is much better for the half-height sample, with $\Delta H=20$ Oe, than the bottom layer structure, with  $\Delta H= 200$ Oe. 

Using the same linewidth values we compared the insertion loss  of the simulations to experimental data in \fig{PyMag}. We see good agreement with the magnitude of loss due to ferromagnetic resonance for the two structures (dotted line). At frequencies far below the ferromagnetic resonance there is a nearly uniform overall offset of the experimental insertion loss from that of the simulation. We suspect that this difference is due to calibration issues in the experiment or possibly roughness in the conductor layers which have not been considered in the simulation.
\section{Conclusion}
We have demonstrated a useful method for characterizing ferromagnetic microstrip structures using the electromagnetic simulation software HFSS.  These simulations provide good agreement with experimental results and can provide a useful tool for investigating these structures. 

\section{Acknowledgements}
This research was supported by US Army Research Office DOA Grant No. W911NF-04-1-0247.
\bibliographystyle{IEEEtran}

%\bibliography{/../../Reference_publications/bibliography/allpapers}
%\bibliography{/Users/jgollub/documents/Reference_Papers/bibliography/allpapers}
%\bibliography{/Users/jgollub/documents/Reference_Papers/bibliography/allpapers}
\end{document}